\RequirePackage{rotating}
\documentclass{aa}  

\usepackage{graphicx}
\usepackage{txfonts}
\begin{document} 

%\graphicspath{/Users/dlennon/Science_esac/gaia}

   \title{Gaia TGAS search for Large Magellanic Cloud runaway supergiant stars}

  \subtitle{Candidate hypervelocity star discovery, and the nature of R\,71}

   \author{Daniel J. Lennon
          \inst{1}
          \and
          Roeland P. van der Marel\inst{2}
          %\fnmsep\thanks{Just to show the usage of the elements in the author field}
          \and
          Mercedes Ramos Lerate\inst{3}
          \and
          William O'Mullane\inst{1}
          \and
          Johannes Sahlmann\inst{1,2}
          }

   \institute{ESA, European Space Astronomy Centre, Apdo. de Correos 78,
E-28691 Villanueva de la Ca\~nada, Madrid, Spain
%             \email{danny.lennon@sciops.esa.int}
         \and
             Space Telescope Science Institute, 3700 San Martin Drive, Baltimore, MD 21218, USA
         \and
            VitrocisetBelgium for ESA, European Space Astronomy Centre, Apdo. de Correos 78,
E-28691 Villanueva de la Ca\~nada, Madrid, Spain
              }

   \date{Received ; accepted}

% \abstract{}{}{}{}{} 
% 5 {} token are mandatory
 
  \abstract
   {}
   {To search for runaway stars in the Large Magellanic Cloud (LMC)
   among the bright {\em Hipparcos} supergiant stars included in the {\em Gaia} DR1 TGAS catalog.}
   {We compute the space velocities of the visually brightest 
   stars in the Large Magellanic Cloud that are
   included in the Gaia TGAS proper motion catalog. 
   This sample of 31 stars contains a Luminous Blue 
   Variable (LBV), emission line stars, blue and yellow supergiants and a SgB[e] star. 
   We combine these results with published radial velocities to 
   derive their space velocities, and by comparing with predictions from stellar dynamical 
   models we obtain their (peculiar)
   velocities relative to their local stellar environment.}
   {Two of the 31 stars have unusually high proper motions. Of the remaining
   29 stars we find that most objects in this sample have velocities that are inconsistent
    with a runaway nature, being in very good agreement with model predictions
    of a circularly rotating disk model.  
    Indeed the excellent fit to the model implies that the TGAS uncertainty
    estimates are likely overestimated. The fastest outliers in this subsample
    contain the LBV R\,71 and a few other well known emission line objects though
    in no case do we derive velocities consistent with fast ($\sim$100\,km/s) 
    runaways. On the contrary our results imply that R\,71 in particular has a moderate 
    deviation from the local stellar velocity field (40\,km/s) lending support to the
    proposition that this object cannot have evolved as a normal single star since it
    lies too far from massive star forming complexes to have arrived at its current position
    during its lifetime. Our findings therefore strengthen the case for this LBV being the
    result of binary evolution. Of the two stars with unusually high proper motions
    we find that one, the isolated B1.5 Ia$^+$ supergiant Sk-67\,2 (HIP\,22237),  is a 
    candidate hypervelocity star, the TGAS proper motion implying a
    very large peculiar transverse velocity ($\sim$360\,km/s) directed radially
    away from the LMC centre.  If confirmed, 
    for example by {\em Gaia} Data Release 2, it would imply that this massive
    supergiant, on the periphery of the LMC, is
    leaving the galaxy where it will explode as a supernova. 
    }
   {}


   \maketitle
%
%-------------------------------------------------------------------

\section{Introduction}

A new model of the stellar dynamics 
of the Large Magellanic Cloud (LMC) was presented by \cite{vdM16} (from
now on vdMS) based on the proper motions of 29
Hipparcos stars that are LMC members and that have suitably precise
proper motions published in the Tycho-Gaia (hereafter TGAS) catalog of 
the first {\em Gaia} Data Release (DR1; \cite{gaiaa, gaiab}, \cite{lindegren}).
Besides their use as test particles of the LMC velocity field, these stars
are of great intrinsic interest due to their representing a sample 
of the visually brightest stars in a nearby star forming galaxy. 
This selection of blue and yellow supergiant stars (see Table 1 and Figure 1) 
includes a number
of well known peculiar and interesting objects such as the
Luminous Blue Variable (LBV) R\,71,  a SgB[e] star (Sk-69\,46),
a candidate LBV (Sk-69\,75; \cite{prinja}) and some peculiar
emission line objects (such as HD37836; \cite{shore}). 

The dynamical properties of these massive stars are of interest.
Those field stars that are relatively isolated, being far from young
clusters, are of particular significance since there are competing
hypotheses to explain their solitude.  Possible explanations include 
{\em in situ} formation of single massive stars (\cite{parker}), or
ejection of runaway stars (with peculiar velocities of order
100 km/s) via disruption of a close binary by a SN explosion 
(\cite{blaauw}) or dynamical interaction in a massive cluster (\cite{poveda}).
Indeed \cite{demink2012, demink2014} point out that binary evolution channels enable 
a continuum of runaway velocities, even including slow runaways, the
so-called walkaways. At the other extreme, even higher peculiar velocities 
(hundreds of km/s) are thought to be generated through interaction with the SMBH at the
centre of our Milky Way (\cite{brown}) producing hyper-velocity stars,
although alternative mechanisms are suggested with the discovery of
hypervelocity B-stars that do not fit this paradigm (\cite{przybilla}). Clearly 
the dynamical signature of these processes is a crucial discriminant in
disentangling these scenarios.

Within the Magellanic Clouds the investigation of runaway stars (in
general) has been limited mainly to circumstantial evidence such as
apparent isolation, peculiar radial velocity, or the presence of
an infrared excess that might be attributed to the presence of a bow 
shock (\cite{evans}, \cite{gvaramadze}). As a specific example we 
point to the current debate as to whether some, or all, LBVs are
runaway stars. \cite{st15} suggests that LBVs are the `kicked mass gainers'
in binary evolution arguing that their relative isolation is inconsistent 
with single star evolution.  Although this hypothesis is heavily debated
(see \cite{hum16}, \cite{smith} and  \cite{davidson}) it is clear that 
the evidence is circumstantial since proper motions for stars in the
LMC cannot yet discriminate between various propositions (but see
\cite{platais} for a first attempt to measure precise relative proper
motions of massive stars around 30 Doradus).  The {\em Gaia} mission
is set to change this picture and fortunately, as noted above,  
our sample contains the most isolated LBV in the LMC, R71,
as well as a number of emission line objects potentially related to LBVs.

In this paper we use the proper
motions from the {\em Gaia} TGAS catalog, along with 
published radial velocities to estimate stellar velocities
relative to their local mean stellar velocity field in order
to search for runaway stars. In Section 2 we describe our methods 
and discuss the results for the sample as whole, while in Section 3
we focus on a few individual objects that are of special 
interest.

 \begin{figure}
   \centering
   \includegraphics[scale=0.6]{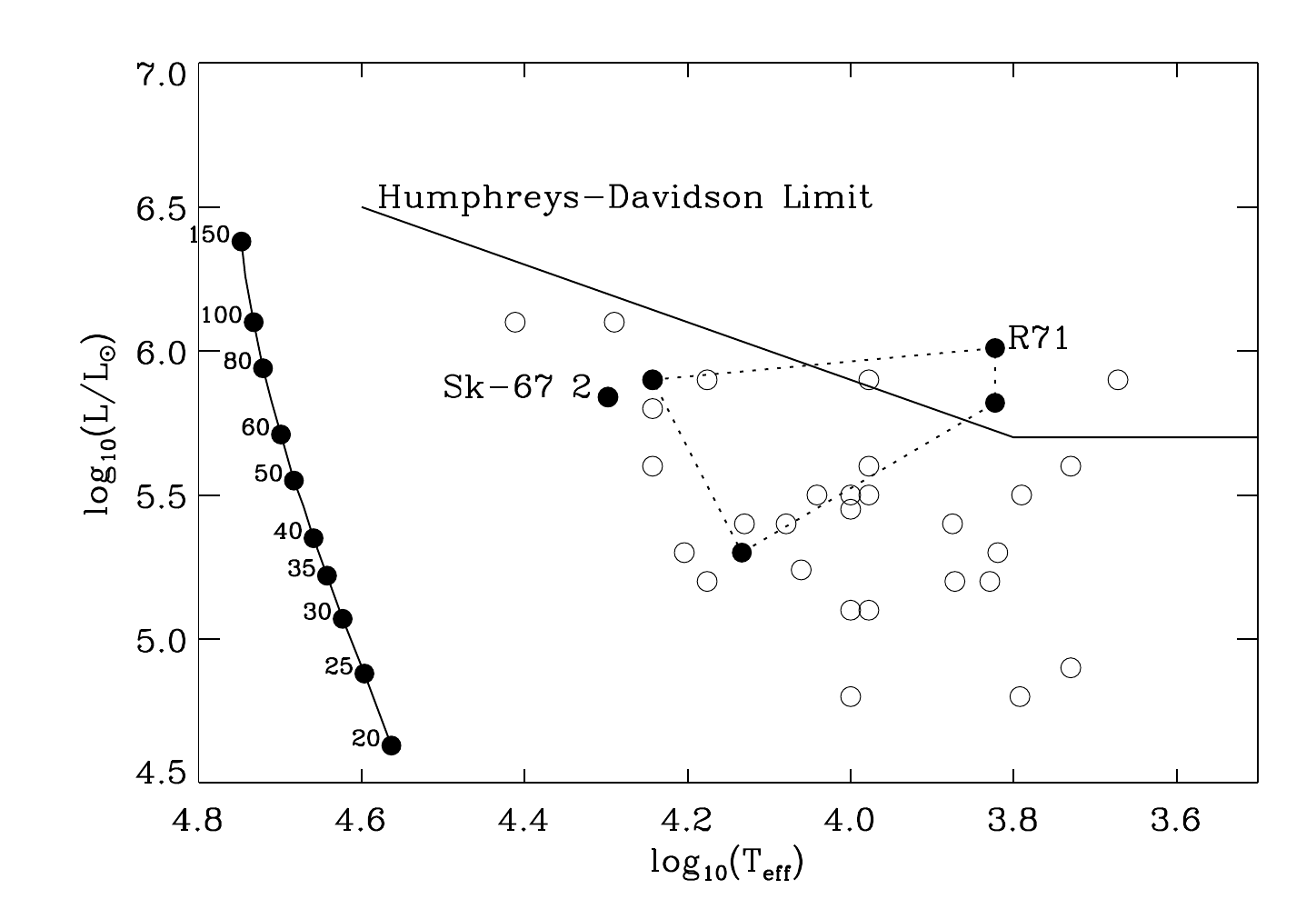}
   \caption{Open symbols represent positions of the stars in our LMC sample in the HR diagram.
   Effective temperatures and luminosities are taken from the literature where available, or
   estimated using spectral types and published photometry. 
   As many of these stars are photometric and spectroscopic variables their positions in the HRD 
   are illustrative. As an extreme example the positions of the LBV R\,71 are indicated (data
   from \cite{mehner}) joined by dotted lines. For convenience we indicate the
   position of the candidate hypervelocity star Sk-67\,2. 
   We also indicate the position of the zero age main sequence (left, labeled with 
   initial masses, tracks from \cite{brott} and \cite{kohler})  and the approximate location of the
   Humphreys-Davidson limit.}
              \label{Fig1}%
    \end{figure}

   \begin{figure}
   \centering
 \includegraphics[scale=0.4]{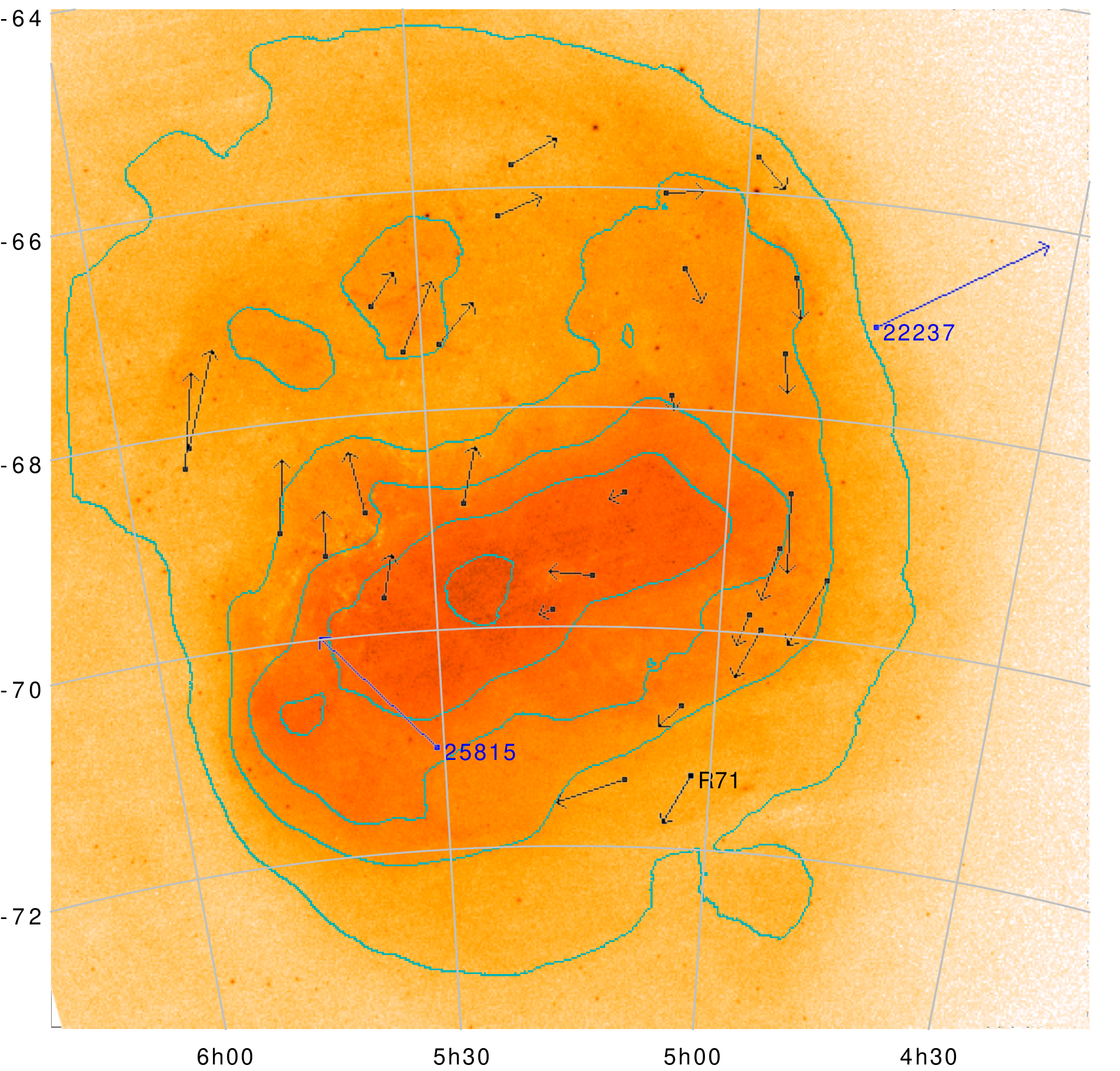}
   \caption{Black arrows indicate magnitudes and directions of proper motions of 
   the 29 stars from vdMS (the LBV R\,71 is labelled), 
   while blue arrows denote positions and proper motions 
   of  Sk-67\,2 (HIP\,22237) and Sk-71\,42 (HIP\,25815) as indicated by the labels. 
   The background image illustrates the stellar density 
   map and contours for the LMC derived from the $Gaia$ catalogue.}
              \label{Fig2}%
    \end{figure}

%--------------------------------------------------------------------
\section{Stellar relative velocities}

Cross-matching the TGAS catalogue with the master catalogue
of LMC massive stars from \cite{bonanos}, supplemented by
the full Sanduleak catalogue (\cite{sanduleak}) of bright LMC members, reveals 
311 sources in common. Examination of the uncertainties in
the measured proper motions reveals two clear groupings; 
31 sources, all  {\em Hipparcos} sources, with proper 
motion uncertainties less than approximately  0.25 mas/yr, and
all others, the {\em TYCHO} sources, with uncertainties in 
excess of approximately 0.6 mas/yr.
In vdMS the TGAS proper motions of 29  of these 31 {\em Hipparcos}
sources 
were used to improve the three dimensional model of the rotation field of 
that galaxy (see also \cite{kroupa}).
This work built upon the previous work of \cite{vdM14}
that modelled this field as a rotating flattened disk, constrained 
by 6790 line-of-sight (LOS) stellar velocities and the average proper
motions of 22 fields in the LMC as derived from {\em Hubble Space Telescope}
observations. Implicit in the use of these 29 stars to constrain this model
is the expectation that their velocities are
a reasonable reflection of the mean stellar motions at their respective
positions. It follows that, say, any runaway star present in this sample,  
with a peculiar relative velocity in excess of around 50\,km/s (about 0.2 mas/yr
in the LMC), would have larger residuals in the model fit
than stars that are not runaway stars. In this context we note that
the formal errors in the proper motions of these stars presented in 
TGAS have a mean value of $\sim$ 0.16 mas/yr in both right ascension and
declination. While this is $\sim$37 km/s in velocity terms 
(see Table 1), as we discuss below we have reason
to believe that the actual errors are substantially smaller than this.
Regardless of this latter point, the precision of the TGAS proper 
motions is even now sufficient to search for (fast) runaway stars
among our sample. 

We therefore compute
the residuals between each star's proper motion, and
the vdMS model prediction, as a measure of its velocity relative to
the mean motion of stars in the LMC. We also compute the 
peculiar LOS velocity by comparing their radial velocities with
the model estimates as constrained by \cite{vdM14} (see their Fig.4).
These results are listed in Table 1 where we have
converted proper motion to velocity assuming a distance to
the LMC of $50.1\pm2.5$ kpc, corresponding to a
distance modulus  of m-M=$18.50\pm0.1$ (\cite{freedman}).
We note that the distance uncertainty also implies a
$\sim$5\% systematic uncertainty in predicted and derived velocities.  
In Tabel 1 we also include two stars, HIP\,22237 (Sk\,-67 2)
and HIP\,25815 (Sk\,-71 42),  that were excluded by vdMS on
the basis that their proper motions are unusual and their 
excess\_astrometric\_noise
parameter is large compared to other stars in the sample.
We have included these two objects in our analysis and 
will return to them in the next section. For now we
point our that their resultant peculiar velocities are 
very large, more akin to hypervelocity stars, and 
we exclude them from the following discussion of the statistical
properties of the sample of 29 stars. However we will return
to these two objects in the next section,  while
the measured proper motions of all 31 stars are illustrated in Fig.2,
that also serves as context for insight into each star's
local environment within the LMC. 

In Fig.\,3 we illustrate our results as velocity dispersion 
histograms in each of the three directions of motion.
One of the striking aspects of the overall results  
is that the goodness of fit to the model is excellent, there are
no outliers beyond 50 km/s in right ascension or declination
(see also Table 1). In fact the standard deviations of the residuals in right ascension and 
declination are each approximately 19 km/s,  
substantially less than the value that might be expected given that 
the formal proper 
motion errors predict a mean of 37\,km/s in velocity 
as mentioned above (well outside the 5\% uncertainty 
in the distance to the LMC).
Fig.\,4 shows the residual vectors in a polar plot indicating that
there is no preferred direction for the peculiar velocities.
We therefore interpret the proper motions of the present sample of stars  as
implying that they do not 
harbour fast runaways (with peculiar velocities in excess of 50\,km/s). 

The LOS velocity information is more difficult to interpret since
measured radial velocities are subject to a number of effects
that introduce off-sets from the true LOS velocity. All of these are 
very luminous (in the approximate range $10^5 -10^6$ solar luminosities) 
and therefore many have strong winds that will effect the centroids of 
strong absorption lines and emission lines. The impact of the former
is small, typically introducing a small blue-shift of $\sim$10\,km/s (\cite{arp}),
while pure emission lines may be even more
blue-shifted since they form in the outflowing wind (see the discussion
of the forbidden [Fe{\,{\sc ii}}] lines of R\,71 in the next section,
and the discussion of the radial velocity of the suspected runaway 
or walkaway VFTS\,682 near 30 Doradus by \cite{vftsIII}).  
The details of course will depend crucially on the parameters of
the stellar wind (stellar radius, mass-loss rate, terminal velocity and velocity law).
Multiplicity can also play a role (it is known that Sk-68\,83 is an eclipsing binary with a
semi-amplitude of $\sim$35\,km/s) and will in general lead to an
overestimate of the velocity dispersion if ignored (\cite{vftsVII}). However
the multiplicity characteristics of these types of stars is poorly
understood at present, and many stars of our sample have only single
epoch measurements.  Given these caveats, and also noting the heterogeneous 
nature of the sources for the radial velocities, it is perhaps not too surprising
to note that the velocity dispersion derived here of $\sim$20 km/s is larger
than the 11.6 km/s derived by \cite{vdM14} for young red supergiants using the same model.

Summarising the results of this section, with the exception of the
two anomalous stars  Sk\,-67 2 and Sk\,-71 42, the proper motions
and LOS velocities of this sample are consistent with the 
hypothesis that none are fast runaways, though there are some
outliers that may be slower runaway stars with peculiar {\em space}
velocities around 50\,km/s. A subset of the 5 fastest slow runaways includes
the isolated B-hypergiant Sk-68\,8, the  LBV R\,71, the candidate
LBV Sk-69\,75, and the SgB[e] star R\,66.

%-------------------------------------- one column figure (place early!)

   \begin{figure}
   \centering
   \includegraphics[scale=0.3]{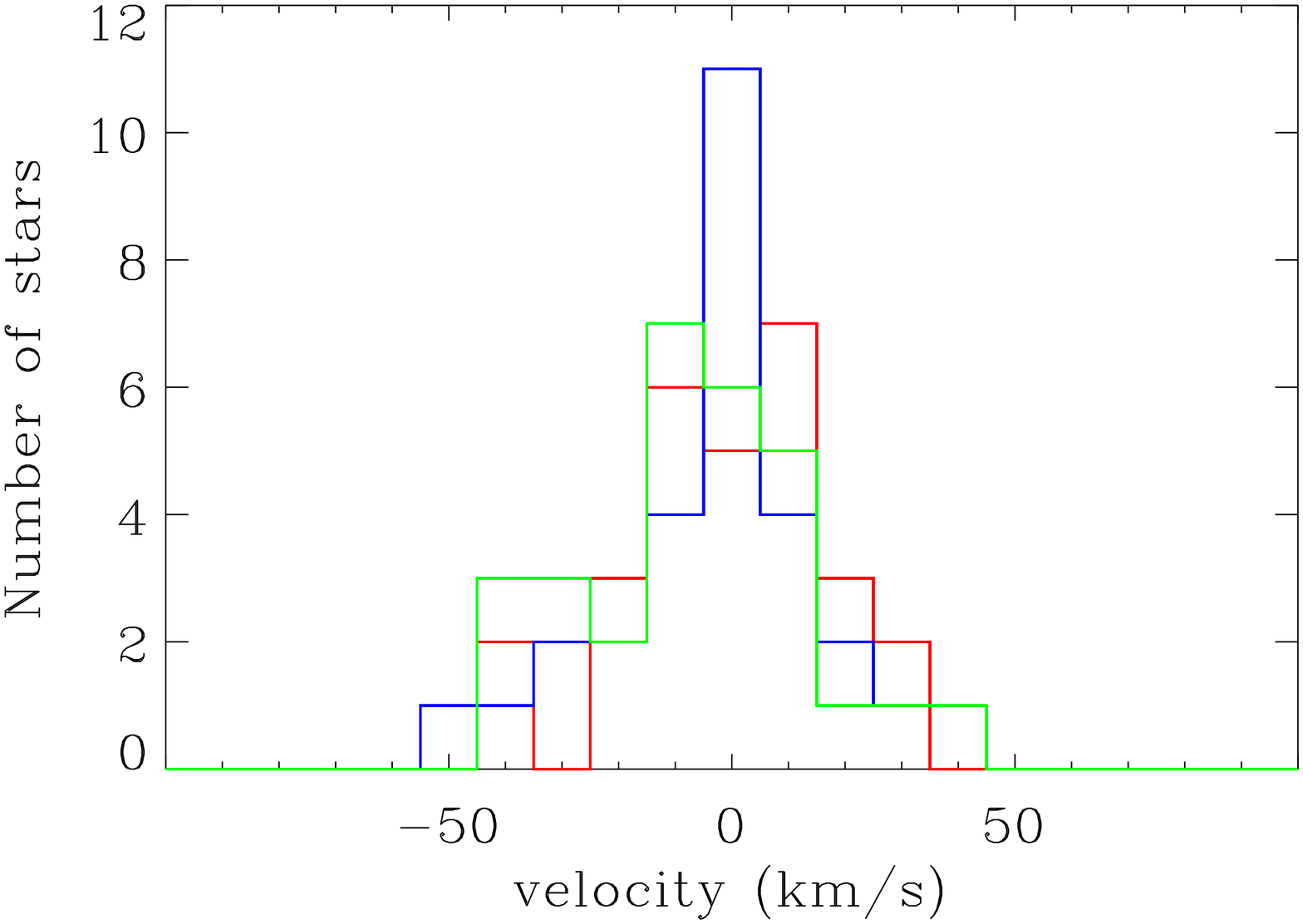}
   \caption{Illustrated here are the velocity residuals in right ascension
   (v$_W$, red), declination (v$_N$, blue) and along the LOS (v$_{LOS}$, green). Histograms have a 
   bin size of 10\,km/s and the mean peculiar 
 velocities in W, N and LOS directions are 2.8, 1.5 and $-0.9$ km/s respectively, with standard 
 deviations of 18.9, 18.8 and 20.3 km/s. Sk-67\,2 (HIP\,22237) and Sk-71\,42 (HIP\,25815) are not
 included in these figures or estimates.}
              \label{Fig3}%
    \end{figure}

   \begin{figure}
   \centering
   \includegraphics[scale=0.55]{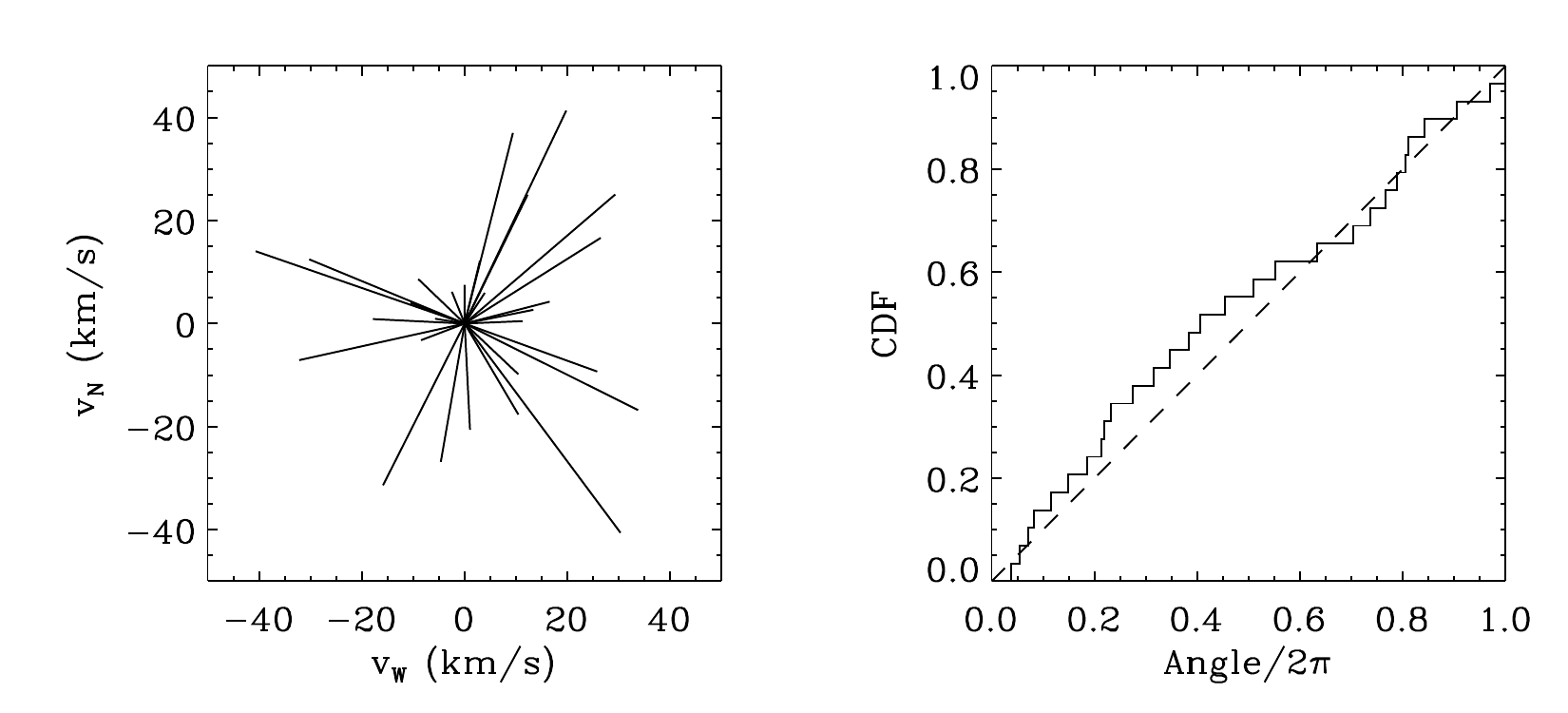}
   \caption{Left panel is a polar plot of  v$_W$ and v$_N$ (from Table 1) while the panel on
   the right illustrates the cumulative distribution function (CDF) of the positions angles of these 
   vectors as a histogram. The straight dashed line represents the CDF of a random 
   distribution.  A formal comparison of the two distributions using the Kuiper test  
   indicates that the observations are consistent with this random distribution
   at a significance level of 85\%.  }
              \label{Fig4}%
    \end{figure}

\section{Discussion: Individual objects of special interest}

In this section we discuss in more detail some implications of our
results for a few objects of special interest.

\subsection{R\,71 (= Sk\,-71 3, HDE\,269006)}

R\,71 is a well studied LBV that has in recent years been undergoing
an outburst,  reaching a maximum visual magnitude  $\sim$8.5 during 
the year 2012.  It is particularly interesting as it is the most isolated LBV in the LMC,
and is also one of the least luminous of this class, but note the influence
of assumptions concerning extinction as pointed out my \cite{mehner},
where more details on this object's recent activity may be found. 

There has been a recent suggestion by \cite{st15} 
that LBVs, such as R\,71, are the rejuvenated mass
gainer products of massive star binary evolution.  Central to this idea is
their proposition that LBVs tend to be found in the field and well away 
from massive star clusters, a consequence of a
more massive star donating material to the initially lower mass
companion, before exploding as a SN and kicking the now more
massive star out of its parent cluster.  This runaway star subsequently
evolves into its LBV phase though now as a relatively isolated
star, before finally exploding as a SN far from its parent cluster.
This is disputed by \cite{hum16} who argue that LBVs are
predominantly the products of single star evolution arguing that the locations of
massive LBVs are well correlated with O-type stars, while the locations of
less luminous LBVs (such as R\,71) correlate well with locations of red supergiants.
They further argue that none of their confirmed LBVs are runaway stars.
In a counter argument \cite{smith} points out that runaway velocities
predicted for the mass-gainer in the binary evolution scenario are not necessarily
high, in many cases ranging from slow (walkaway) velocities of a few km/s to
a few tens of km/s (\cite{demink2014}).
The rejuvenated star also has a lifetime longer than
would be expected for one of its luminosity if it were a single star,
aiding and abetting the appearance of unusual isolation.
Interestingly they attribute a LOS peculiar velocity of $-71$\,km/s
to R\,71 arguing that it is consistent with it being a potential fast runaway, but see our
discussion below, and the argument in \cite{davidson} concerning
this star's LOS velocity.  While we cannot shed light on 
the LBV population of the LMC as a whole we can address the nature of  R\,71.
As discussed by \cite{st15} it is  the most isolated
of the confirmed LBVs in the LMC. They estimate that there is not
any O-star within 300\,pc, the closest being almost half a degree
distant (or 450\,pc) in projection. 

Concerning R\,71's LOS velocity, we adopt a value of
204\,km/s (see Table 1), compared to the value of 192\,km/s used by 
\cite{mehner} that is commonly also used by other authors. 
As they report however
this latter estimate is based on the [Fe\,{\sc ii}] lines. 
However \cite{wobig} reports a discrepancy between these lines and the
absorption lines of neutral helium, metals and higher series Balmer lines 
that imply a systemic radial velocity of 204$\pm 5$\,kms, a result also obtained
by \cite{wolf81}. Since the [Fe\,{\sc ii}] lines form in the expanding wind 
(as also noted by both \cite{mehner} and \cite{wolf81}), we adopt
the higher radial velocity of 204\,km/s that is weighted more towards the weaker
photospheric absorption features in its spectrum that should be less influenced by
its wind. We caution however that
even these lines may well be slightly impacted and the actual
systemic LOS velocity may be a little higher.   Comparing with
the model predicted LOS velocity at this position of 238\,km/s gives a peculiar
LOS velocity of only $-34$\,km/s. (Comparing to the average  
red and yellow supergiant velocities 
from \cite{neugent} produces a similar result, implying a mean 
velocity at this position of $\sim$235 km/s and a very slightly lower
LOS peculiar velocity of $-31$ km/s).

Our results for R\,71 indicate that both its proper motion and LOS
velocity are most likely inconsistent with it being a fast runaway star
(with peculiar velocity in the range 50--100\,km/s), having a transverse velocity 
of around 38\,km/s (see Table 1) though with a formal uncertainty
of $\pm 34$\,km/s (but note our assessment that more realistic
errors may be roughly half this value).  At this velocity it would take almost 
12\,Myr to travel the 450\,pc from the nearest grouping of OB stars,
or about 6\,Myr assuming the uncertainty adds to the velocity (or 
9\,Myr with our more optimistic error estimate). These numbers
are difficult to reconcile with single-star main-sequence lifetimes of
approximately 8, 5, 4 and 3 Myrs for stars with initial masses of 
20, 30, 40 and 60\,M$_{\odot}$ (\cite{brott}). 

\subsection{Sk-67\,2 (= HDE\,270754 = HIP\,22237)}

Due to its moderately high extinction of $E(B-V)$$\sim$~$0.2$,  Sk\,-67\,2 has been the subject
of many observations with various facilities as a means of studying the interstellar
medium of the LMC (see \cite{welty} for an example). 
The star is also classified by \cite{fitz} as belonging
to a those B-type supergiants, denoted as {\em N weak} or {\em BC},
that appear nitrogen deficient with respect to 
the morphologically normal nitrogen rich supergiants.
We find that this isolated star (it lies at the
north-west edge of the galaxy, see Fig.2) 
has a very large peculiar velocity of 359\,km/s (Table 1)
in a direction away from the LMC. 
%Taken at face value this
%implies that Sk-67\,2 is a hypervelocity star, but of course we note
%that vdMS omitted this object from their sample, citing its
%astrometric\_excess\_noise parameter being greater than their
%cut-off value of 1.0. In fact the value of this 
%parameter is 1.11, and is not an obvious outlier compared to
%the 29 stars that vdMS used in their solution. 

This star was omitted from the vdMS sample, in part because its peculiar
proper motion implies that it cannot help constrain the overall LMC
rotation field. However, they also noted that this star has the
second-highest astrometric\_excess\_noise parameter $a$ in the sample,
after Sk-71 42 (see Section 3.3). However, while the latter star ($a$ =
2.58) is truly a strong outlier, this is not the case for Sk-67 2 ($a$
= 1.11). The remaining 29 sample stars have a mean $\langle a \rangle
= 0.51$ with an RMS of $0.18$. There are two other stars in the sample
with $a \sim 0.9$, HIP 22849 and 27868, and these do not have unusual
proper motions (see Table 1). Therefore, the astrometric\_excess\_noise
for Sk-67 2 does not necessarily imply that its measured TGAS proper
motion must suffer from unidentified systematics. It is consistent
with being an $\sim$3$\sigma$ random Gaussian outlier in the overall
distribution.  We also note that Sk-67 2 has a strong mid- and far-infrared
excess (\cite{bonanos}; see also Fig.5) that
would be also be consistent with a high peculiar velocity if
that motion were to drive a bow shock as the star ploughs through
the interstellar medium.  

%However Trundle (private communication; in preparation) reports that
%these BC and {\em N weak} stars may exhibit evidence for a 
%mild oxygen enhancement ($\sim$0.2\,dex) with respect to normal supergiants.

 We have a high resolution {\sc feros} echelle spectrum of this object
 obtained in 2004, and we have extracted high resolution VLT/UVES
 echelle spectra from the ESO archive obtained in 2001 and 2012. 
 Radial velocities measured 
 from the weak metal lines in these data lead to a radial velocity of
 $320\pm1$\,km/s for all three epochs.
 Comparing our measured radial velocity with
the predicted mean LOS velocity from \cite{vdM14} at this position of 277\,km/s 
leads to a peculiar velocity of $+43$\,km/s (comparing to the average red and yellow supergiant 
velocities from \cite{neugent} lead to a similar peculiar
LOS velocity of $+49$\,km/s). \cite{vdM14} also cite a velocity dispersion for the
young massive stars in the LMC of 11.6\,km/s implying that Sk-67\,2 is 
a clear $\sim$4$\sigma$ outlier, though perhaps not such an obvious outlier in LOS as in
proper motion. (Using instead $\sigma$=20 km/s from Fig.3 implies it is
a 2.3$\sigma$ outlier.)

Based on the above evidence we argue that this star is a true LMC member
and not some peculiar foreground Halo object and, since its velocity vector
is almost parallel to the disk plane (see below), it is highly unlikely that it is a 
more distant star of unexplained origin or nature.
To understand whether or not Sk-67 2 is bound we must compare
its velocity with the escape velocity of the LMC. While this is
currently uncertain, we can obtain an approximate value by comparison
with a Milky Way value of 533 km/s as derived from the 
RAVE survey (\cite{piffl}). The ratio of these galaxies'  total dark 
halo masses of 0.061 (\cite{guo}) then implies an escape velocity 
for the LMC of $\sim131$ km/s. Alternatively if we directly compare circular 
velocities for the LMC (92 km/s, \cite{vdM14}) and Milky Way
(239 km/s, \cite{mcmillan}) we obtain an escape velocity of
around 205 km/s. Therefore Sk-67 2 is clearly unbound with respect to the
LMC and, as we argue below, a possible hypervelocity star.

While Sk-67 2 is unbound its velocity is still significantly below the
surface escape velocity of main sequence star of its approximate 
mass and therefore not an unambiguous candidate for a 
hypervelocity star since there are theoretical scenarios in which
fast runaway stars, or hyper-runaway stars, might be created
by classical runaway star scenarios. However, $N$-body simulations of
dynamical ejection from massive clusters (\cite{perets}) predict
very low rates of production of hyper-runaway stars, while
binary evolution models have difficulty producing any runaway stars
with velocities in excess of $\sim200$ km/s (\cite{eldridge}).  Based on
current models therefore it would appear that the hyper-runaway 
scenario is not a promising explanation of the nature of Sk-67 2. 

One can calculate the 3D position and 3D velocity vectors of Sk-67\,2 
in an LMC-centric frame  using the approach used in
\cite{vdM2002} (this is a one-parameter family
depending on the star's actual distance). 
The minimum angle between its position and velocity vectors 
in this frame is 5.3$\pm$5.2 degrees, attained for
a distance that is 50.6 kpc (i.e., 0.5 kpc larger than the distance of
the LMC, but 0.5 kpc in front of the inclined LMC disk at this
location). The uncertainty in this angle is dominated by the
uncertainty in the position of the LMC center ($\sim 0.3$ deg per
coordinate), which was taken from the joint fit to the HST and TGAS PM
data in vdMS.
In other words, its velocity vector is almost aligned with the
LMC centre, as would be expected for a hypervelocity star ejected 
by a central massive black hole.  
We note here however
that the star is $\sim 3.5$ kpc from the centre of the LMC
implying that if the star distance is such that it is $\pm$3.5 kpc from the
LMC disk plane, then the angle between the velocity vector and the
vector to the LMC centre would be $\sim$45 degrees.

While no such black hole is known to exist, the presence of a black hole 
of mass $\leq 10^{7}$M$_{\odot}$ cannot
be ruled out by the observed velocity field (Boyce et al.\,(2017), in preparation).
Interestingly \cite{boubert} have suggested that the presence of
a central massive black hole  in the LMC might naturally explain 
the clustering of known hypervelocity stars in the constellations Leo and Sextens,
provided that there is an as yet undiscovered southern population loosely 
focused on the LMC. 

Given that its lifetime (below) clearly rules out a Milky Way origin, 
it is tempting to attribute membership of this population of
hypervelocity stars to Sk-67 2. However
there are still some issues with this explanation. For example, 
its distance from the centre of the LMC is
$\sim$3.5 kpc that would take $\sim$10 Myr to traverse at its current
velocity. Even allowing for the fact that we have ignored the 
effect of the LMC gravitational potential in estimating its flight time,
implying this is an upper limit, it is still more than
a factor of two longer than its likely main sequence lifetime. If, like R\,71, this
star is a product of binary evolution, having been a hypervelocity
binary ejected from a massive black hole at the centre of the LMC,
then indeed there might be evolutionary channels that could be
consistent with its current position. On the other hand, this discrepancy
between lifetime and flight-time may provide some support for
the hypothesis that some hypervelocity stars originate via 
interaction with an intermediate mass black hole (IMBH) as
proposed by  \cite{przybilla} for the
star HE\,0437-5439, thought to have originated from the LMC.
However the LMC is not known to host a confirmed IMBH either.

Stars in the LMC with unusual line-of-sight kinematics were previously
reported by \cite{olsen}. These stars have velocity residuals
up to $\sim 100$ km/s, and lower metallicities than typical for the
LMC. They were interpreted as a population of accreted SMC stars. The
three-dimensional velocity difference between the LMC and SMC is $128
\pm 32$ km/s (\cite{kallivayalil}), consistent with the scale of
the observed residual velocities for these stars. By contrast, Sk-67 2
has a velocity residual of $\sim 360$ km/s compared to the
LMC. Moreover, direct comparison to the known three-dimensional
velocity of the SMC implies a residual velocity of $\sim 336$ km/s
with respect to that galaxy. Therefore, Sk-67 2 cannot have an SMC
origin (which would also be hard to reconcile with its short main
sequence lifetime).

Finally in this subsection we note that the TGAS proper motion is essentially 
derived from the difference between
the {\em Hipparcos} and {\em Gaia} positions of this object (since its
parallax is negligible). One might suppose that some perturbation of 
the point-spread-function, such as the presence of a close visual companion, 
might cause an error in the estimated position. However Sk-67\,2 is in a sparsely 
populated part of the LMC, and our high resolution {\sc feros}
spectrum shows no evidence for a bright companion. Further, to cause a
spurious proper motion of $\sim$1\,mas/yr in TGAS,  one would 
require an offset of  $\sim$20\,mas in the Gaia astrometry. This 
is of order one-fifth of an AL/AC-average pixel and in addition this offset 
would have to be consistent for the different scan angles, that seems rather unlikely.
In this context we also note that very good agreement is found between proper motions 
derived from Hubble and TGAS for the Magellanic Clouds (vdMS) and 
Globular Clusters in the Milky Way (\cite{wat}).
We conclude that {\em Sk-67\,2 is a candidate hypervelocity star} that if confirmed, by
for example {\em Gaia} Data Release 2 (DR2), would warrant further investigation
as to the nature of its origin.

   \begin{figure}
   \centering
   \includegraphics[scale=0.6]{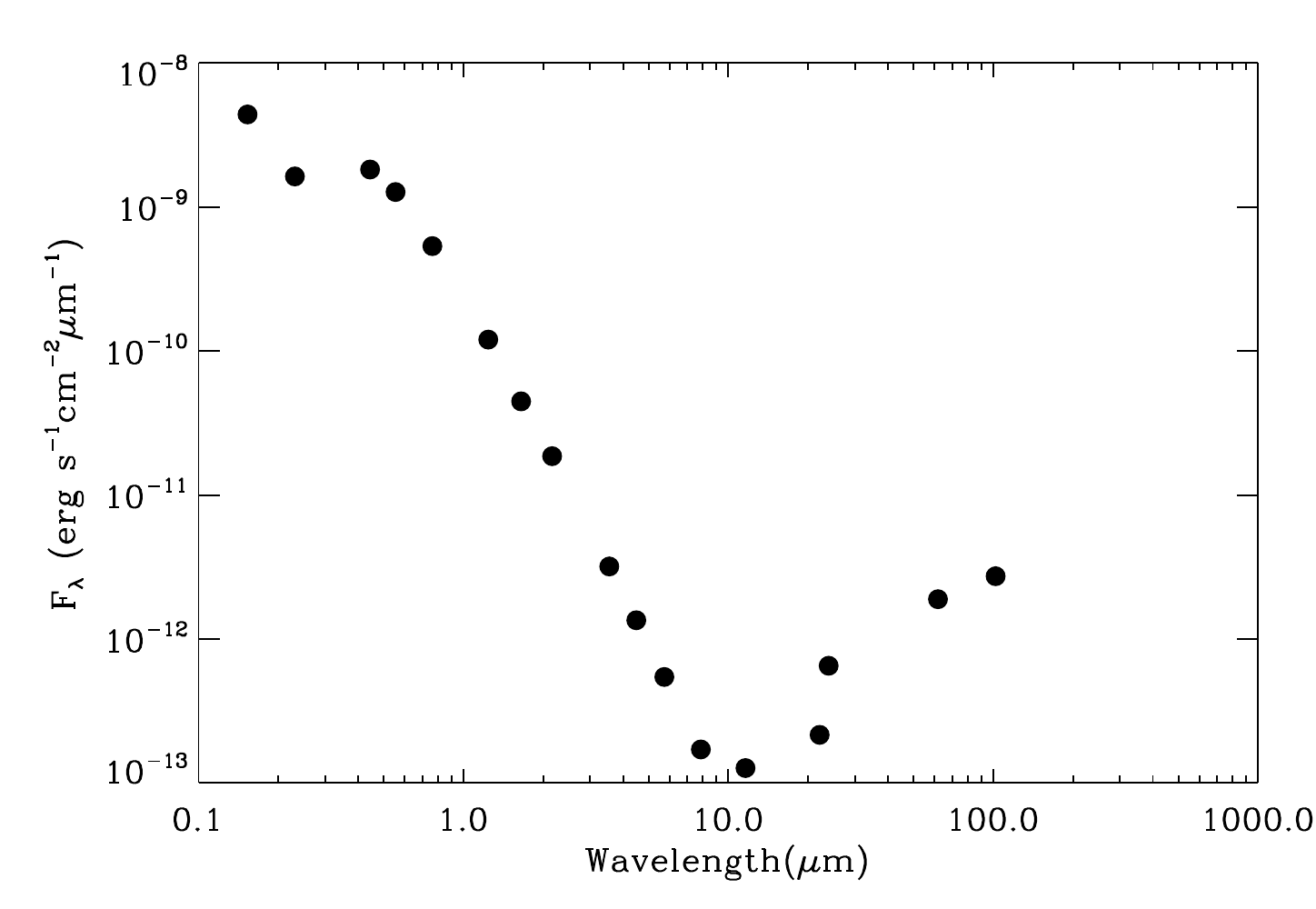}
   \caption{Spectral Energy Distribution of the candidate hypervelocity star
   Sk-67\,2 showing its infrared excess from
   about 10 to 100 $\mu$m that is a signature of dust emission and a possible
   bow-shock (based on {\em Spitzer}, {\em WISE} and $IRAS$ data).}
              \label{Fig5}%
    \end{figure}

\subsection{Sk-71\,42 (=HDE\,269660 = HIP\,25815)}

The TGAS proper motions imply a peculiar velocity of $\sim$150 km/s 
for Sk-71\,42 but
in contrast to Sk-67\,2, it is in a very crowded
region of the LMC and lies (in projection) within the environment  
of the nebula and supernova remnant LHA 120-N206 that hosts
many massive O-type stars. Also, by contrast, this star has
astrometric\_excess\_noise=2.58, substantially above the values
of the rest of the present sample. Also, 
comparing our measured radial velocity of 238\,km/s with
the mean LOS velocity predicted by the model
of \cite{vdM14} at this position of 239\,km/s  gives a peculiar
LOS velocity of only $-1$\,km/s, while from the monitoring of 
\cite{morrell} we know that this star has constant radial velocity.
(Comparing to the average red and yellow supergiant velocities  
of \cite{neugent}
yields a peculiar velocity of $-12$\,km/s.)
The lack of a peculiar LOS velocity, together with its unusually large
astrometric\_excess\_noise parameter suggests that we must await 
further data on this object from
{\em Gaia} DR2 before speculating further on its nature.

\section{Summary}

We have shown that the {\em Gaia} TGAS catalogue is even now,
with DR1, able to provide important dynamical constraints on a
subset of the visually brightest massive stars in the LMC that can
be used to address their nature as potential walkaway, runaway
or hypervelocity stars. Specific conclusions concerning this
sample include;
\begin{itemize}
\item Most of these very luminous stars are not runaways, the outliers
mostly have peculiar velocities of less than 50\,km/s.
\item R\,71 is rather unique in its isolation and we have shown that it's
peculiar space velocity is only moderately discrepant from its
environment, that is difficult to reconcile with a single-star
evolutionary scenario. A potential solution to this dilemma might be
that this particular LBV is indeed the result of binary evolution
but is the evolved product of a slow runaway binary (either a 
rejuvenated mass gainer or a stellar merger). 
\item The isolated B1.5\,Ia$^+$ supergiant Sk-67\,2 is found to be
a candidate hypervelocity star with a peculiar velocity of $\sim$360
km/s directed radially away from the LMC centre, suggesting possible
ejection by an as yet undiscovered central black hole. However its main sequence
lifetime is difficult to reconcile with the likely flight-time suggesting
alternative hypotheses for the origin of its velocity will need to be explored.
\end{itemize}

\begin{acknowledgements}
This work has
made use of data from the ESA space mission Gaia
(http://www.cosmos.esa.int/gaia), processed by the
Gaia Data Processing and Analysis Consortium (DPAC,
http://www.cosmos.esa.int/web/gaia/dpac/consortium).
Funding for the DPAC has been provided by national
institutions, in particular the institutions participating
in the Gaia Multilateral Agreement. DJL thanks Hassan Siddiqui,
Uwe Lammers and Jose Hernandez of the ESAC {\em Gaia} Science
Operations Centre for useful discussions. This research made use
of Simbad and Vizier provided by CDS, Strasbourg;  ESASky, developed by the ESAC
 Science Data Centre (ESDC); the ESO archive at Garching; and
 TOPCAT. J.\,Sahlmann was supported by an ESA
Research Fellowship in Space Science. 

\end{acknowledgements}


%\input{table_data_v3.tex}

\clearpage
\onecolumn

\begin{sidewaystable}
\centering
\caption{Basic information for programme stars together with velocities (in km/s) relative to the vdMS model as follows; v$_W$ and v$_N$ are velocities
in (minus) right ascension (West) and declination (North), e$_W$ and e$_N$ are the errors of these values (from TGAS). We also list the peculiar transverse velocity (or speed) on the sky,
v$_{\rm t}$, as the quadrature sum
of these velocities and e$_{\rm t}$ the combined error assuming their errors are uncorrelated. v$_{\rm rad}$ is the measured radial velocity and
v$_{LOS} $ = v$_{\rm rad} - {\rm v}_{\rm rad(model)}$ is the relative velocity in the line of sight relative to the prediction of the \cite{vdM14} rotating disk model.}
%\centering
\begin{tabular}{llrlrlrllrlll}
\hline\hline
Name & HIP & v$_W$ & e$_W$ & v$_N$ & e$_N$ & v$_{\rm t}$ & e$_{\rm t}$ & v$_{\rm rad}$ &  v$_{LOS}$	& Common aliases & Sp.Type\tablefootmark{b} & $Gaia$ source Id.   \\
\hline		
Sk-67 2\tablefootmark{a}  & 22237 & 251.4 & 27.8 & 256.7 & 24.6 & 359.3 & 26.2 & 320  &    43.0  & R\,51, HDE\,270754 	& B1.5 Ia$^+$   	& 4662413602288962560\\
Sk-69 7   &  	22392 & 	-8.5	 & 35.8	 & -3.2	 & 35.8	 & 9.1	 & 35.8	 & 263	   & 4.3   & R\,52, HDE\,268654 	& B9 Ia         	& 4655349652394811136\\
Sk-68 8  & 	22758 & 	30.3	 & 37.9	 & -40.6 & 40.6	 & 50.7	 & 39.7	 & 271	   & -1.3 & R\,58, HDE\,268729 	& B5 Ia$^+$    		& 4655510043652327552\\
Sk-69 30 & 	22794 & 	11.2	 & 20.8	 & 0.5	 & 20.9	 & 11.3	 & 20.8	 & 258	   & -8.1  & R\,59, HDE\,268757 	& G7 Ia$^+$     	& 4655460771785226880\\
Sk-66 12 & 	22849 & 	-9.0	 & 28.6	 & 8.6	 & 30.8	 & 12.5	 & 29.7	 & 284	   & -4.4  & R\,61, HDE\,268675 	& B8 Ia         	& 4661769941306044416\\
Sk-67 19 & 	22885 & 	3.0	 & 34.5	 & 12.2	 & 35.0	 & 12.6	 & 35.0	 & 304	   & 17.6  & HDE\,268719          	& A2 Ia         	& 4661720532007512320\\
Sk-69 37 & 	22900 & 	13.4	 & 55.1	 & 2.7	 & 52.5	 & 13.6	 & 55.0	 & 252	   & -3.0  & HDE\,268819          	& F5 Ia          	& 4655136518933846784\\
R\,66    & 	22989 & 	29.3	 & 55.7	 & 25.1	 & 53.9	 & 38.6	 & 54.9	 & 264	   &  6.4 & Sk-69 46, HDE\,268835, S\,73 & SgB[e]         	& 4655158131209278464\\
Sk-65 8	 & 	23177 & 	-2.5	 & 38.4	 & 6.2	 & 37.7	 & 6.7	 & 37.8	 & 311	   & 15.3  & HDE\,270920 		& G0 Ia          	& 4662293892954562048\\
R\,71	 & 	23428 & 	33.7	 & 34.7	 & -16.8 & 33.4	 & 37.7	 & 34.4	 & 204	   & -33.5 & Sk-71\,3, HDE\,269006, S\,155 & LBV       	& 4654621500815442816\\
Sk-70 48 & 	23527 & 	26.5	 & 57.1	 & 16.6	 & 54.7	 & 31.3	 & 56.4	 & 216	   & -28.3 & HDE\,269018, GV\,183 	& B2.5 Ia         	& 4655036841335115392\\
Sk-66 58 & 	23665 & 	1.0	 & 24.8	 & -20.6 & 27.2	 & 20.7	 & 27.2	 & 300	   & -0.2  & R\,75, HDE\,268946 	& A0 Ia           	& 4661920986713556352\\
Sk-67 44 & 	23718 & 	0.0	 & 17.1	 & 7.5	 & 18.8	 & 7.5	 & 18.8	 & 255	   & -37.4  & R\,76, HD33579, S\,171 	&A2 Ia$^+$       	& 4661472145451256576\\
HDE\,271018 & 	23820 & 	12.2	 & 50.4	 & 25.0	 & 51.3	 & 27.8	 & 51.2	 & 291	   & -11.5  & GV\,538 			& F8 Ia           	& 4662061311885050624\\
Sk-71 14 & 	24006 & 	-32.2	 & 24.9	 & -7.1	 & 23.3	 & 33.0	 & 24.9	 & 243	   & -7.9 & R\,80, HDE\,269172 	& A0 Ia          	& 4651629489160555392\\
Sk-68 63 & 	24080 & 	-10.5	 & 21.9	 & 4.0	 & 22.4	 & 11.3	 & 21.9	 & 253	   & -24.9  & R\,81, HDE\,269128, S\,86 & B2.5 Ia$^+$ 	& 4658269336800428672\\
Sk-69 75 & 	24347 & 	-40.7	 & 46.7	 & 14.0	 & 42.0	 & 43.0	 & 46.2	 & 240	   & -25.2 & HDE\,269216, S\,88 	& B8 I      		& 4658204053297963392\\
Sk-69 91 & 	24694 & 	10.4	 & 31.1	 & -17.7 & 29.9	 & 20.5	 & 30.2	 & 230	   & -30.2 & HDE\,269327, S\,95 	& B2 Iae        	& 4658137739001073280\\
Sk-65 48 & 	24988 & 	3.9	 & 18.7	 & 6.0	 & 21.1	 & 7.1	 & 20.4	 & 322	   & 12.7  & R\,92, HDE\,271182 	& F8 Ia    		& 4660601607121368704\\
Sk-66 72 & 	25097 & 	10.4	 & 41.1	 & -9.8	 & 45.5	 & 14.4	 & 43.2	 & 305	   & -3.8  & HDE\,269408	 	& A0 Ia    		& 4660444926713007872\\
Sk-68 81 & 	25448 & 	19.8	 & 32.8	 & 41.3	 & 31.7	 & 45.8	 & 31.9	 & 276	   & -1.0  & HDE\,269541, W\,9-26 	& A2 Ia      		& 4658486455992620416\\
Sk-67 125 & 	25615 & 	16.5	 & 48.6	 & 4.3	 & 49.3	 & 17.1	 & 48.6	 & 320	 & 18.2  & HDE\,269594 & 		G0 Ia        		& 4660175580731856128\\
Sk-71 42\tablefootmark{a}  & 	25815 & -110.3  & 28.5 & 103.1 & 30.2 & 150.9 & 29.3 & 238 & -1.0 & R\,112, HDE\,269660 	& B2 Ia       		& 4651835303997699712\\
Sk-67 162 & 	25892 & 	9.4	 & 43.3	 & 37.0	 & 44.2	 & 38.2	 & 44.2	 & 311	   & 10.4  & HDE\,269697, W\,24-21	& F6 Ia        		& 4660124762671796096\\
Sk-67 201 & 	26135 & 	-15.9	 & 22.7	 & -31.4 & 26.9	 & 35.2	 & 26.1	 & 342	   & 38.7  & R\,118, HDE\,269781 	& A0 Iae      		& 4660246224352015232\\
Sk-69 201 & 	26222 & 	25.8	 & 44.5	 & -9.3	 & 44.6	 & 27.4	 & 44.5	 & 267	   & 4.1   & R\,123, HD\,37836, S\,124,MWC\,121 & B0 Iae  	& 4657280635327480832\\
Sk-68 131 & 	26338 & 	-30.3	 & 44.0	 & 12.5	 & 47.5	 & 32.7	 & 44.5	 & 276	   & 0.0  & HDE\,269857, W\,27-61 	& A9 Ia       		& 4657700408260606592\\
Sk-69 267 & 	26745 & 	-4.6	 & 54.8	 & -26.8 & 57.4	 & 27.2	 & 57.4	 & 250	   & -19.6 & R\,151W, HDE\,269982, W\,8-37 & A8 Ia        	& 4657627943562907520\\
Sk-69 299 & 	27142 & 	-5.8	 & 33.7	 & 0.9	 & 32.5	 & 5.9	 & 33.7	 & 275	   & 0.6  & R\,153, HDE\,270086, S\,143 & A2 Ia$^+$      	& 4657722879521554176\\
Sk-68 178 & 	27819 & 	2.1	 & 36.4	 & 0.4	 & 25.2	 & 2.2	 & 36.1	 & 313	   & 23.4 & HDE\,270296 		& B6 Ia           	& 4659188769038018816\\
Sk-68 179 & 	27868 & 	-17.9	 & 36.7	 & 0.9	 & 26.3	 & 17.9	 & 36.7	 & 333	   & 45.7  & HDE\,270305 		& B3 Ia        		& 4659091084305723392\\
\\
Mean\tablefootmark{c}     && 2.8  & 37.0 & 1.5  & 36.6 & 22.7 & & & -0.9 && \\
$\sigma$\tablefootmark{c} && 18.9 &      & 18.8 &      & 13.7 & & & 20.3 && \\
\hline
\end{tabular}
\tablefoot{
\tablefoottext{a}{These two stars were excluded from vdMS on the basis of their outlying proper motions and excess\_astrometric\_noise parameter.}
\tablefoottext{b}{Spectral types are from the literature excepting those of Sk-69\,91, that was derived from its low resolution IUE spectrum, and 
Sk-67\,125, derived from its broad band colour (and is hence uncertain).}
\tablefoottext{c}{Excluding Sk\,-67 2 and Sk\,-71 42.}
}
\end{sidewaystable}

\twocolumn
\end{document}